\documentclass[letterpaper]{article}

\usepackage{url,txfonts}
\usepackage{apacite,graphicx}

\title{\bf Experiential AI}
\author{
{\bf Drew Hemment}, {University of Edinburgh} \quad {\tt drew.hemment@ed.ac.uk} \\   
{\bf Ruth Aylett}, {Heriot-Watt University}  \quad {\tt r.s.aylett@hw.ac.uk} \\ 
{\bf Vaishak Belle}, {University of Edinburgh}  \quad {\tt Vaishak@ed.ac.uk} \\ 
{\bf Dave Murray-Rust}, {University of Edinburgh}  \quad {\tt D.Murray-Rust@ed.ac.uk} \\ 
{\bf Ewa Luger}, {University of Edinburgh} \quad {\tt Ewa.Luger@ed.ac.uk} \\ 
{\bf Jane Hillston}, {University of Edinburgh}\quad  {\tt Jane.Hillston@ed.ac.uk} \\ 
{\bf Michael Rovatsos}, {University of Edinburgh} \quad  {\tt Michael.Rovatsos@ed.ac.uk} \\ 
{\bf Frank Broz}, {Heriot-Watt University} \quad  {\tt f.broz@hw.ac.uk}
}
\date{}

\begin{document}
\maketitle
	



 

 \begin{abstract}  \it 
     Experiential AI is proposed as a new research agenda in which artists and scientists come together to dispel the mystery of algorithms and make their mechanisms vividly apparent. It addresses the challenge of finding novel ways of opening up the field of artificial intelligence to greater transparency and collaboration between human and machine. The hypothesis is that art can mediate between computer code and human comprehension to overcome the limitations of explanations in and for AI systems. Artists can make the boundaries of systems visible and offer novel ways to make the reasoning of AI transparent and decipherable. Beyond this, artistic practice can explore new configurations of humans and algorithms, mapping the terrain of inter-agencies between people and machines. This helps to viscerally understand the complex causal chains in environments with AI components, including questions about what data to collect or who to collect it about, how the algorithms are chosen, commissioned and configured or how humans are conditioned by their participation in algorithmic processes.

 \end{abstract}
\section{Introduction}
AI has once again become a major topic of conversation for policy makers in industrial nations and a large section of the public. 

In 2017, the UK published Ready, Willing and Able, a landscape report \cite{house2018ai}. It clearly states that "everyone must have access to the opportunities provided by AI" and argues the need for public understanding of, and engagement with AI to develop alongside innovations in the field. The report warns of the very real risk of "societal and regional inequalities emerging as a consequence of the adoption of AI and advances in automation" ((\textit{Ibid}.). It also assesses issues of possible harm from malfunctioning AI, and resulting legal liabilities. However, it stops short of considering more pervasive downsides of applying AI decision-making across society. Alongside the sometimes exaggerated claims of AI’s current or immediate-future capabilities, a broader set of fears about negative social consequences arise from the fast-paced deployment of AI technologies and a misplaced sense of trust in automated recommendations. While some of these fears may themselves be exaggerated, negative outcomes of ill-designed data-driven machine learning technologies are apparent, for example where new knowledge is formulated on undesirably biased training sets. The notorious case of Google Photos grouping some humans with primates on the basis of skin tone offered a glimpse of the damage that can be done. Such outcomes may not be limited to recommendations on a mobile phone: social robots share everyday spaces with humans, and might also be trained on impoverished datasets. Imagine, for example, a driverless car not recognizing specific humans as objects it must not crash into. So much for Asimov's laws!

\maketitle
 
\section{Accountability and explainability in AI}

The AI community has, of course, not been silent on these issues, and a broad range of solutions have been proposed. We broadly classify these efforts into two related categories: accountability and explainability. 

The first category seeks to identify the technical themes that would make AI trustworthy and accountable. Indeed, we can see AI technologies are already extending the domains of automated decision making into areas where we currently rely on sensitive human judgements. This raises a fundamental issue of democratic accountability, since challenging an automated decision often results in the response 'it’s what the computer says'. So operators of AI need to know the limits and bounds of the system, the way that bias may present in the training data, or we will see more prejudice amplified and translated to inequality. From the viewpoint of AI research, there is a growing scientific literature on fairness \cite{kleinberg2018algorithmic} to protect those otherwise disenfranchised through algorithmic decisions, as well as engineering efforts to expose the limitations of systems. Accountability can be a deeper property of the system too: for example, an emerging area of AI research looks at how ethical AI systems might be designed  \cite{conitzer2017moral,halpern2018towards,hammond2018deep}. 

The second category investigates how the decisions and actions of machines can be made explicable to human users \cite{gunning2017explainable}. We are seeing a step change in the number of people both currently and potentially impacted by automated decisions. Whilst the use of algorithms can now be said to be common \cite{domingos2015master}, concerns arise where complex systems are applied in the generation of sensitive social judgments, such as in social welfare, healthcare, criminal justice, and education. his has led to a call to limit the use of 'black box' systems in such settings \cite{campolo2017ai}. However, if one asks for a rationale for a decision, usually none is given, not least because those working in organisations using automated decision-making do not themselves have any insight into what the algorithms driving it are doing. This is a form of conditioning, creating passivity rather than engagement. At the other extreme, if people do not understand the decisions of AI systems, they may simply not use those systems. Be that as it may, progress in the field has been exciting but a single solution is elusive. Some strands of research focus on using simpler models (possibly at the cost of prediction accuracy), others attempt "local" explanations that identify interpretable patterns in regions of interest \cite{weld2018intelligible,ribeiro2016should}, while still others attempt human-readable reconstructions of high-dimensional data \cite{penkov2017using,belle2017logic}. However, this work addresses explainability as primarily a technical problem, and does not account for human, legal, regulatory or institutional factors. What is more, it does not generate the kind of explanations needed from a human point of view. A person will want to know why there was one decision and not another, the causal chain, not an opaque description of machine logic. There are distinctions to be explored between artificial and augmented intelligences \cite{carter2017using}, and a science, and an art, to be developed around human-centred machine learning \cite{fiebrink2018HumanCentred}.

For there to be responsible AI, transparency is vital, and people need comprehensible explanations. Core to this is the notion that unless the operation of a system is visible, and people can access comprehensible explanations, it cannot be held to account. However, even when explanation can be provided, it may not always be sufficient \cite{edwards2017slave}. There is a need for more intuitive interventions to, for example, integrate domain knowledge in ways that connect managers with those at the frontlines, or understand the changing relations between data and the world \cite{veale2018fairness}. In Seeing without knowing, Ananny and Crawford argue research needs not to look \textit{within} a technical system, but to look \textit{across} systems and to address both human and non-human dimensions \cite{ananny2018seeing}. We propose that art offers one way to answer their call for ``a deeper engagement with the material and ideological realities of contemporary computation"  (\textit{Ibid}.). 

\begin{figure*}
  \includegraphics[width=\textwidth]{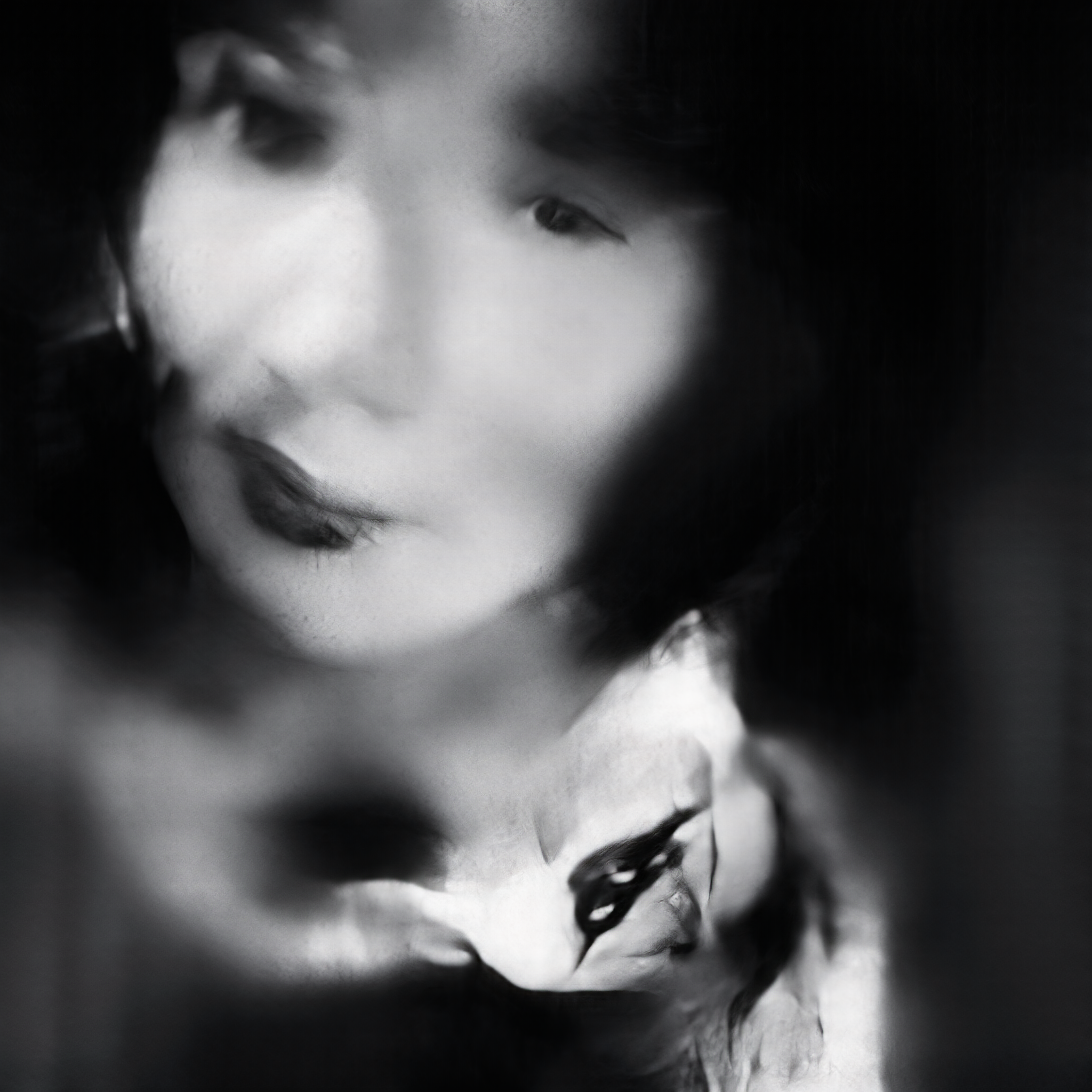}
  \caption{Neural Glitch 1540737325 $\copyright$  Mario Klingemann 2018}
  \label{klingermann}
\end{figure*}


\section{Artists addressing such AI challenges}

There is a mature tradition of work between art and technology innovation going back to the 1960s and 1970s \cite{harris1999xerox, gere2009digital}. Artists are beginning to experiment in AI as subject and tool, and several high profile programmes are testament to the fertility of this field \cite
{encodingCultures,mediaArtBetween}. Such practice can create experiences around social impacts and consequences of technology, and create insights to feed into the design of the technologies \cite{hemment2017art}. 

One theme evident among artists working with machine learning algorithms today, such as Mario Klingemann\footnote{\url{http://quasimondo.com/}} and Robbie Barrat\footnote{\url{https://robbiebarrat.github.io/}}, is to reveal distortions in the ways algorithms make sense of the world -- see Figure \ref{klingermann} for an example. This kind of approach enables the character of machine reasoning and vision to be made explicit, and its artifacts to be made tangible. This in turn creates a concrete artefact or representation that can be used as an object for discussion and to spark further enquiry, helping to build literacy in those systems. 

In the contemporary experience of AI,
the disturbing yet compelling output of
DeepDream has shaped our view on what
algorithms do, although it is
questionable how representative this is
of deep network structures, or whether
it is a happy accident in machine
aesthetics. Either way, it has prompted
artistic exploration of the social
implications of AI, with projects using
deep learning to generate faces
\cite{reddit2019portraits} and
Christies auctioning neural network
generated portraits
\cite{christies2019portrait}. Going
beyond the typical human+computer view,
artists are questioning the
construction of prejudice and normalcy \cite{mushon},
and working with AI
driven prosthetics, to open
possibilities for more intimate
entanglements
\cite{donnarumma2019AIProsthetics}.

Art can both make ethical standards concrete, and allow us to imagine other realities. While high-level ethical principles are easy to articulate, they sit at a level of generality that may make their practical requirements less obvious. Equally, they signal the existence of clear solutions, externalise responsibility, and obscure the true complexity of the moral problems resulting from socially situated AI. Ethical issues must be concretely internalised by developers and users alike to avoid failures like Cambridge Analytics or the Facebook Emotional Contagion experiment \cite{jouhki2016facebook}. Experiential approaches \cite{kolb2014experiential} can act as a powerful mechanism, and embedding relevant experiences in a story-world through narrative, and especially role-play, can generate safe reflection spaces – as for example Boal's Forum Theatre \cite{boal2013rainbow}. 

Accountability is variously addressed. Joy Buolamwini works with verse and code to challenge harmful bias in AI\footnote{\url{https://www.poetofcode.com/}}, while Trevor Paglen constructs a set of rules for algorithmic systems in such a way as to uncover the character of that rule space\footnote{\url{http://www.paglen.com/}}. A thriving community of practitioners from across the arts and sciences are working to avoid detection \footnote{\url{https://cvdazzle.com/}} or trick classification systems \cite{sharif2016accessorize}. Such artistic experiments brings to life and question what an algorithm does, what a system could be used for, and who is in control.

\maketitle
 
\section{Experiential AI theme and call for artists}

The field of Experiential AI seeks to engage practitioners in computation, science, art and design around an exploration of how humans and artificial intelligences relate, through the physical and digital worlds, through decisions and shaping behaviour, through collaboration and co-creation, through intervening in existing situations and through creating new configurations. 

The Experiential AI theme begins with a call for artists in residence, launched in August 2019, as a collaboration between the Experiential AI group at University of Edinburgh, Ars Electronica in Linz, and Edinburgh International Festival \footnote{\url{https://efi.ed.ac.uk/art-and-ai-artist-residency-and-research-programme-announced/}}. The focus is on creative experiments in which AI scientists and artists are jointly engaged to make artificial intelligence and machine learning tangible, interpretable, and accessible to the intervention of a user or audience. The ambition is help us think differently about how algorithms should be designed, and open possibilities for radically new concepts and paradigms.




\end{document}